# Spectral CT Reconstruction via Low-rank Representation and Structure Preserving Regularization

Yuanwei He[1,2], Li Zeng[1,2], Qiong Xu[3,4], Zhe Wang[3,4], Haijun Yu[2,5], Zhaoqiang Shen[1,2], Zhaojun Yang[1,2], Rifeng Zhou[2,5,6]

[1] College of Mathematics and Statistics, Chongqing University, Chongqing 401331, China
[2] Engineering Research Center of Industrial Computed Tomography Nondestructive Testing, Ministry of Education, Chongqing University, Chongqing 400044, China
[3] Beijing Engineering Research Center of Radiographic Techniques and Equipment, Institute of High Energy Physics, Chinese Academy of Sciences, Beijing 100049, China
[4] Jinan Laboratory of Applied Nuclear Science, Jinan 250131, China
[5] Key Lab of Optoelectronic Technology and Systems, Ministry of Education, Chongqing University, Chongqing 400044, China
[6] State Key Laboratory of Mechanical Transmission, Chongqing University, Chongqing 400044, China

**Correspondence**
Li Zeng, College of Mathematics and Statistics, Chongqing University, Chongqing 401331, China.
Email: drlizeng@ cqu.edu.cn

**Abstract**
**Objective:** With the development of computed tomography (CT) imaging technology, it is possible to acquire multi-energy data by spectral CT. Being different from conventional CT, the X-ray energy spectrum of spectral CT is cutting into several narrow bins which leads to the result that only a part of photon can be collected in each individual energy channel, which cause the image qualities to be severely degraded by noise and artifacts. To address this problem, we propose a spectral CT reconstruction algorithm based on low-rank representation and structure preserving regularization in this paper.
**Approach:** To make full use of the prior knowledge about both the inter-channel correlation and the sparsity in gradient domain of inner-channel data, this paper combines a low-rank correlation descriptor with a structure extraction operator as priori regularization terms for spectral CT reconstruction. Furthermore, a split-Bregman based iterative algorithm is developed to solve the reconstruction model. Finally, we propose a multi-channel adaptive parameters generation strategy according to CT values of each individual energy channel.
**Main results:** Experimental results on numerical simulations and real mouse data indicate that the proposed algorithm achieves higher accuracy on both reconstruction and material decomposition than the methods based on simultaneous algebraic reconstruction technique (SART), total variation minimization (TVM), total variation with low-rank (LRTV), and spatial-spectral cube matching frame (SSCMF). Compared with SART, our algorithm improves the feature similarity (FSIM) by 40.4% on average for numerical simulation reconstruction, whereas TVM, LRTV, and SSCMF correspond to 26.1%, 28.2%, and 29.5%, respectively.
**Significance:** We outline a multi-channel reconstruction algorithm tailored for spectral CT. The qualitative and quantitative comparisons present a significant improvement of image quality, indicating its promising potential in spectral CT imaging.

## 1. INTRODUCTION

X-ray computed tomography (CT), as a noninvasive technology, is widely exploited in medical procedures, industrial nondestructive testing, security check, and geosciences[1-4]. Though conventional CT has achieved a great success during the past decades, it has some inevitable limitations on identification and classification of different tissues. Conventional CT systems do not discriminate between photon energies. The obtained CT images present the average attenuation of X-rays, while failing to describe the spectrum distribution of X-ray photons. The CT values of different substances are similar in the circumstances. It means that tissues composed by different elements are indistinguishable according to conventional CT images. As a matter of fact, attenuation characteristics of X-ray vary with energies, which reflects the physical properties of inspected materials. More information about the tissue composition can be presented by analyzing the energy information of X-ray.

Spectral CT is proposed according to the extension of conventional CT along the energy dimension. Though spectral CT dates from the same era as CT, clinical and industrial implementations of spectral CT lagged behind the initial prediction due to the limits of technology[5]. With the development of CT imaging technology and the increase of clinical requirements, spectral CT imaging has come to the fore over the past decade. The rapid tube potential switching, multilayer detectors, and multi X-ray sources make it possible to acquire multi-energy data by spectral CT, so that the energy information in polychromatic X-ray can be utilized for optimizing tissue characterization[6-9]. Compared with conventional CT, spectral CT can significantly improve diagnostic performance in lesion detection, material decomposition, differential diagnosis, and tissue characterization[10-12]. However, spectral CT only collects a part of photons into each individual energy channel which leads to the result that the reconstructed images are severely degraded by noise and artifacts.

In order to reconstruct high quality images from spectral CT data, a number of reconstruction models are proposed recently. Some reconstruction methods for low-dose CT can be used for spectral CT as well, including total variation (TV), total generalized variation, non-local means, dictionary learning, wavelet, and so on[13-16]. With the combination of TV, Xu et al. proposed a compressive sensing-based statistical interior tomography algorithm for the reconstruction of each energy bin image[14]. By penalizing the higher-order derivatives of images, Zeng et al. developed penalized weighted least-squares scheme for spectral CT image reconstruction[17]. However, the models mentioned above only reconstruct images independently on individual energy channel, while ignoring the relationship among different channels. In order to take advantages of inter-channel correlations, prior information based on the spectral direction of spectral CT is integrated into reconstruction process. The non-local image similarity among different energy channels is one of the mainstream prior information for spectral CT reconstruction. Li et al. exploited spatio-spectral features by introducing non-local means to make use of the redundant information in spectral CT[18]. Niu et al. collected patches from multi-energy images and utilized the self-similarity of them for spectral CT reconstruction[19]. Yao et al. took advantages of image correlations along both spatial and spectral directions to suppress noise during spectral CT reconstruction process[20]. Hu et al. proposed an enhanced-sparsity reconstruction method, utilizing the non-local feature similarity in the spatial-spectral domain by clustering similar spatial-spectral patches within non-local window to a 4th-order tensor group[21]. Wu et al. developed a spatial-spectral cube matching frame (SSCMF) to make full use of the spatial-spectral space similarity for spectral CT reconstruction[22]. Low-rank prior, as an advanced prior using for the strong synergy among different energy channels, has also been widely applied in spectral CT reconstruction. Chu et al. proposed a method integrating TV and the trace norm of image tensor as sparsity and low-rank constraints for multi-energy CT reconstruction[23]. Kim et al. proposed a penalized maximum likelihood method using spectral patch-based low-rank penalty[24]. Shi et al. combined a region-specific texture model with a low-rank correlation descriptor as an a priori regularization to explore a superior texture preserving Bayesian reconstruction for spectral CT[25].

In this paper, we attempt to take full use of the prior knowledge about both the sparsity in gradient domain of inner-channel data and the inter-channel correlations to construct a reconstruction model for spectral CT. Considering the characteristics of severe noise and artifacts in each individual energy channel, we first propose bilateral weighted relative total variation (BRTV) to simultaneously extract main structures and fine details as well as further restrain noise and artifacts for image restoration. Being different from relative total variation (RTV)[26], the weights in BRTV rely on both the closeness to vicinity in the domain and the similarity to vicinity in the range, which could better characterize the distance and difference between two pixels. Two close and similar pixels make less contributions to BRTV. Minimizing BRTV could suppress noise while preserve sharp edges and fine details at the same time. Then we propose a new spectral CT reconstruction model, combining a low-rank correlation descriptor and a structure extraction operator as prior regularization terms. The nuclear norm and BRTV are respectively utilized to characterize the inter-channel correlations and extract the inner-channel structures. This model can not only inherit the advantages of low-rank correlation descriptor in image feature recovery, but also eliminate noise-induced artifacts in flat area by the proposed BRTV regularizer. A corresponding split-Bregman based iterative algorithm is developed to solve the optimization problem mentioned above. Furthermore, to address the selection of parameters in different energy bins, we propose a multi-channel adaptive parameters selection strategy in terms of CT values in each individual energy channel to dynamically generate parameters corresponding to different channels. We need only empirically select parameters for the first channel, and the parameters for other energy channels can be adaptively generated.

In summary, the main contribution of this work can be summarized as fourfold. Firstly, given the prior knowledge about sparsity in gradient domain of spectral CT inner-channel data, we propose an image restoration operator, BRTV, which is based on the proximity and similarity between two pixels in a local rectangular. Secondly, we present a spectral CT reconstruction model, in which the nuclear norm and BRTV are respectively utilized as a low-rank correlation descriptor and a structure extraction operator to constrain the inter-channel correlations and extract the inner-channel structures. Thirdly, an efficient iterative algorithm based on split-Bregman is designed for solving this model[27]. Finally, we propose an adaptive parameters generation strategy to dynamically generate parameters for different energy channels.

The rest of this paper is structurally organized as follows. In section 2, we briefly review the previous knowledge in regard to our model, and give an introduction of our BRTV and the reconstruction model. Then we present a multi-channel adaptive parameters generation strategy used for our reconstruction algorithm. In section 3, the experimental results of the numerical simulations and real mouse datasets are demonstrated. Finally, the conclusion of this work is given.

## 2. METHOD

In this section, we first give a brief review on low-rank recovery for spectral CT imaging, and mathematically present the spectral CT model. Then, we propose the bilateral weighted relative total variation (BRTV) and the spectral CT reconstruction model. Furthermore, we develop an iterative algorithm for solving this model. A multi-channel adaptive parameters generation strategy corresponding to this algorithm is presented at the end of this section.

### 2.1 Low-rank recovery for spectral CT imaging

For spectral CT, images in different energy channels describe the same object so that images among different energy channels are highly redundant. Previous studies indicate that low-rank matrices can be recovered from most of the sampled subsets[25,28]. It means that spectral CT images can be sufficiently characterized by their low-rank approximation. In this paper, the concept of tensor is used to build links among images in different energy channels, and we use nuclear norm acting on a spectral CT image tensor to approximate low-rank. The spectral CT images are represented as a three dimensional tensor.

A $S$ dimensional tensor is defined as $\mathcal{X} \in \mathcal{R}^{N_1 \times N_2 \times \ldots \times N_S}$. To compute the rank of a tensor, we can use the mode-$s$ unfolding operator to reshape the tensor into a matrix first.

$$unfold_s(\mathcal{X}) := \mathcal{X}_{(s)} \in \mathcal{R}^{N_s \times (N_1 \ldots N_{s-1} N_{s+1} \ldots N_S)}. \tag{1}$$

It is worth noting that the tensor can be recovered by a folding operator.

$$fold_s(\mathcal{X}_{(s)}) := \mathcal{X} \tag{2}$$

By turning the tensor into a matrix, the rank of a tensor can be approximated by the sum-weighted nuclear norm of all its mode-$s$ unfolding matrices[29], which is presented as:

$$Rank(\mathcal{X}) = \|\mathcal{X}\|_* = \sum_{s=1}^{S} \alpha_s \|\mathcal{X}_{(s)}\|_*, \tag{3}$$

where $\sum_{s=1}^{S} \alpha_s = 1$, $\|\mathcal{X}_{(s)}\|_*$ is the nuclear norm of matrix $\mathcal{X}_{(s)}$ with the definition as follows:

$$\|\mathcal{X}_{(s)}\|_* = \sum_m \sigma_m(\mathcal{X}_{(s)}), \tag{4}$$

where $\sigma_m(\mathcal{X}_{(s)})$ is the $m^{th}$ singular value of $\mathcal{X}_{(s)}$.

In spectral CT, images in different energy channels are highly correlated, hence we only utilize low-rank to constrain the mode-3 unfolding matrix of the three dimensional tensor. Therefore, we set $\alpha_1 = \alpha_2 = 0$ in this paper[25].

## 2.2 Spectral CT images reconstruction model

Taking the data noise into account, the conventional forward model for CT imaging can be considered as a linear system:

$$Y = AX + \eta, \tag{5}$$

where $Y \in \mathcal{R}^J (J = J_1 \times J_2)$ represents the vectorized projection data degraded by noise $\eta \in \mathcal{R}^J$. $J_1$ and $J_2$ are the view and detector numbers, respectively. $X \in \mathcal{R}^{N_I} (N_I = N_1 \times N_2)$ is the vectorized two-dimensional image. $N_1$ and $N_2$ are the width and height of the reconstructed image. $A \in \mathcal{R}^{J \times N_I}$ stands for the CT system matrix.

Results of the problem shown in Eq. (5) usually cannot be directly obtained by the inverse matrix due to the limitation of computer memory. In general, it can be solved by minimizing the following objective function:

$$\arg\min_X \|AX - Y\|_2^2, \tag{6}$$

where $\|\cdot\|_2$ is the $L_2$ norm. ART and SART are the typical methods to solve the above problem. Due to the presence of noise in the projection data, a regularization term is generally integrated to stabilize the solution:

$$\arg\min_X \|AX - Y\|_2^2 + \lambda R(X). \tag{7}$$

The first term in Eq. (7) is a data fidelity term. $R(X)$ is the regularization term. $\lambda$ denotes the penalty parameter to balance the data fidelity term and the regularization term.

For spectral CT, multiple projection data are produced at different energy ranges[30]. Detectors collect these datasets of the same imaging object with one scan. Each energy-dependent image can be reconstructed by each projection data, respectively. The spectral CT reconstruction model can be constructed as:

$$\arg\min_{\mathcal{X}} \sum_{s=1}^{S} \|AX_s - Y_s\|_2^2 + \lambda R(\mathcal{X}), \tag{8}$$

where $X_s \in \mathcal{R}^{N_I} (N_I = N_1 \times N_2)$ and $Y_s \in \mathcal{R}^J (J = J_1 \times J_2)$ are the vectorized image and the projection of the $s^{th} (s = 1, 2, ..., S)$ energy channel, respectively. $\mathcal{X} \in \mathcal{R}^{N_1 \times N_2 \times S}$ is the reconstructed image tensor.

## 2.3 Bilateral weighted relative total variation

Recently, relative total variation (RTV) was proposed to extract main structures from images with complex textures[26]. In other words, textures can be smoothed by minimizing RTV. According to the definition of RTV, it is easy to confuse noise and textures[31], which means that noise and textures tend to be smoothed at the same time. As a result, fuzzy edges and textures are represented in restored results by RTV especially for spectral CT images which suffer from high-level noise. We consider that this phenomenon is related to the weighting function of RTV, which is only based on the distance between pixels, while the differences between image gray-scale values are neglected. To address this issue, we draw lessons from the thought of bilateral filtering[32,33] to utilize spatial proximity and pixel value similarity as a portion of the weighting function in RTV so that details can be distinguished from noise. Therefore, bilateral weighted relative total variation (BRTV) is proposed to simultaneously maintain edge and further reduce noise. BRTV can be expressed as:

$$BRTV(p) = \frac{D_x(p)}{\widetilde{L}_x(p) + \varepsilon} + \frac{D_y(p)}{\widetilde{L}_y(p) + \varepsilon}, \tag{9}$$

where $\varepsilon$ is a small number to avoid denominators equaling to zero. $D_x(p)$ and $D_y(p)$ are called windowed total variations (WTV) along $x$ and $y$ direction at the $p^{th}$ pixel of image $X$, using for reflecting the absolute differences between neighbouring pixels within a rectangular region $R(p)$ centered at $p^{th}$ pixel. Structures, textures and noise in region $R(p)$ contribute to WTV, which has no dependence on the sign of each partial derivative. $\widetilde{L}_x(p)$, and $\widetilde{L}_y(p)$ are the modified windowed inherent variations (mWIV) along $x$ and $y$ direction at the $p^{th}$ pixel of image $X$. The weight in mWIV is different from that in RTV. The mWIV, integrating a composite similarity measure $h_{p,q}$, helps to distinguish details and noise in image $X$. WTV and mWIV are defined as:

$$D_x(p) = \sum_{q \in R(p)} k_{p,q} \cdot \left|(\partial_x X)_q\right|, \quad D_y(p) = \sum_{q \in R(p)} k_{p,q} \cdot \left|(\partial_y X)_q\right|, \tag{10}$$

$$\widetilde{L}_x(p) = \left|\sum_{q \in R(p)} h_{p,q} \cdot (\partial_x X)_q\right|, \quad \widetilde{L}_y(p) = \left|\sum_{q \in R(p)} h_{p,q} \cdot (\partial_y X)_q\right|, \tag{11}$$

where $\partial_x$ and $\partial_y$ are the partial derivatives in two directions. $k_{p,q}$, defined based on the closeness between the $p^{th}$ and the $q^{th}$ pixels, is the original weighting function in RTV. The $q^{th}$ pixel belongs to region $R(p)$. $k_{p,q}$ was expressed as:

$$k_{p,q} \propto \exp\left(-\frac{\|p-q\|^2}{2\sigma^2}\right), \tag{12}$$

where $\|p-q\|$ is the Euclidean distance between the $p^{th}$ and the $q^{th}$ pixels. $\sigma$ relates to the scale of window $R(p)$.

$h_{p,q}$ is proportional to the product of position proximity and gray similarity between the $p^{th}$ and the $q^{th}$ pixels:

$$h_{p,q} \propto \exp\left(-\frac{\|p-q\|^2}{2\sigma^2}\right) \cdot \exp\left(-\frac{\|X(p)-X(q)\|^2}{2\sigma^2}\right), \tag{13}$$

where $X(p)$ and $X(p)$ are grayscale values at the $p^{th}$ and the $q^{th}$ pixels. In the scalar case, $\|X(p)-X(q)\|$ represents the absolute difference of the grayscale values. A large $h_{p,q}$ represents a greater closeness referring to distance and gray level between the $p^{th}$ and the $q^{th}$ pixels. As a result of that, the partial derivatives make more contributions for mWIV to reach a smaller BRTV.

**2.4 The proposed spectral CT reconstruction model**
In this section, we propose a new model for spectral CT reconstruction by combining a low-rank correlation descriptor and a structure extraction operator as prior regularization terms to characterize the inter-channel correlations and the inner-channel data sparsity in gradient domain. Considering the two dimensional fan-beam scan geometry, the proposed spectral CT reconstruction model is set as:

$$\arg\min_{\mathcal{X}} \sum_{s=1}^{S} \left( \|AX_s - Y_s\|_2^2 + \lambda \sum_p BRTV(X_s(p)) \right) + \mu \|\mathcal{X}\|_*, \tag{14}$$

where $X_s(p)$ represents the $p^{th}$ pixel in image $X_s$. $\|\mathcal{X}\|_*$ denotes a nuclear norm of the reconstructed image tensor $\mathcal{X}$. $\lambda$ and $\mu$ are parameters set to balance the data fidelity term and regularization terms.

By using the augmented Lagrangian method, Eq. (14) can be rewritten as the following unconstrained optimization problem:

$$\arg\min_{\mathcal{X},\mathcal{F},\mathcal{V}} \sum_{s=1}^{S} \left( \|AX_s - Y_s\|_2^2 + \lambda \sum_p BRTV(X_s(p)) \right) + \mu \|\mathcal{F}\|_* + \beta \|\mathcal{X} - \mathcal{F} - \mathcal{V}\|_F^2, \tag{15}$$

where $\mathcal{F} \in \mathcal{R}^{N_1 \times N_2 \times S}$ is an auxiliary variable introduced to represent $\mathcal{X}$. $\mathcal{V} \in \mathcal{R}^{N_1 \times N_2 \times S}$ denotes the error feedback variable. $\beta$ is the Lagrangian parameter. According to the split-Bregman method[27], each variable can be updated with the following sub-problems.

**Sub-problem 1:**

$$\mathcal{X}^{k+1} = \arg\min_{\mathcal{X}} \sum_{s=1}^{S} \left( \|AX_s - Y_s\|_2^2 + \lambda \sum_p BRTV(X_s(p)) \right) + \beta \|\mathcal{X} - \mathcal{F}^k - \mathcal{V}^k\|_F^2, \tag{16-a}$$

**Sub-problem 2:**

$$\mathcal{F}^{k+1} = \arg\min_{\mathcal{F}} \mu \|\mathcal{F}\|_* + \beta \|\mathcal{X}^{k+1} - \mathcal{F} - \mathcal{V}^k\|_F^2, \tag{16-b}$$

**Sub-problem 3:**

$$\mathcal{V}^{k+1} = \mathcal{V}^k + \mathcal{F}^{k+1} - \mathcal{X}^{k+1}. \tag{16-c}$$

In the above, $k+1$ denotes the current iteration step. More description about sub-problem 1

and sub-problem 2 are given in Appendix A and B, respectively. The sub-problem 3 can be calculated directly.

The complete workflow of the proposed spectral CT reconstruction algorithm is summarized as follows.

---
**WORKFLOW FOR THE PROPOSED MODEL**

**Initialization:** $\mathcal{X}$, $\mathcal{F}$, and $\mathcal{V}$;

**Input:** $\beta$, $\lambda$, $\mu$, $\sigma$, $\varepsilon_{X_s}$, and $N_{BRTV}$.

**While stop criterion is not met:**

    Step 1: Update $\mathcal{X}^{k+1}$ by Eq. (16-a);
    Step 2: Update $\mathcal{F}^{k+1}$ by Eq. (16-b);
    Step 3: Update $\mathcal{V}^{k+1}$ by Eq. (16-c);
    $k \leftarrow k+1$

**End until the stop criterion is satisfied.**

---

## 2.5 Multi-channel adaptive parameters

It is worth noting that the noise level in different energy channels is different. Using fixed regularization parameters for BRTV in all of the energy channels, called multi-channel consistent parameters selection strategy, tends to cause over-smoothing, edge blurring, and detail missing. Parameters need to vary from channel to channel due to the differences of data among energy channels. Therefore, a multi-channel adaptive parameters generation strategy is proposed in terms of CT values in each individual energy channel to dynamically generate parameters for different channels. We need only select parameters for the first channel, and parameters for other energy channels can be adaptively generated.

Regularization parameter $\lambda$ in Eq. (14) is used to control the BRTV weight in each energy channel. An appropriate parameter $\lambda$ makes contribution to minimize the objective function value and obtain an outstanding performance on noise suppression. BRTV is severely penalized with a large $\lambda$, which means that noise in images can be greatly suppressed. In other words, Parameter $\lambda$ should depend on the CT values of each energy channel. Specifically, a CT image with larger CT values should correspond to a greater penalty intensity. In our model, $\lambda$ is set to be proportional to the sum of CT values in each individual energy channel.

## 3. Experiments

In this section, experiments on numerical simulation and real mouse datasets are employed to validate and evaluate the performance of the proposed algorithm. Simultaneous algebraic reconstruction technique (SART)[34], total variation minimization (TVM)[14], total variation with low-rank (LRTV)[35], and spatial-spectral cube matching frame (SSCMF)[22] are used for comparison. All of the experiments ran in the MATLAB (R2016a) environment on a desktop equipped with NVIDIA GeForce RTX 3060. In numerical simulation experiments, we demonstrate the performance of different algorithms from both evaluation indexes and visual effects. Root mean square error (RMSE), peak signal-to-noise ratio (PSNR), and feature similarity (FSIM) are used to quantitatively analyse. Regarding real mouse experiments, excellent image quality and accurate decomposition results are considered for parameter optimization. In addition, post-reconstruction material decomposition is carried out to evaluate the reconstruction performance of different methods.

In this paper, we tuned BRTV related parameters ($\lambda$, $\sigma$, $\varepsilon$, and $N_{BRTV}$) rely on the experience from Gong et al. [31], in which they gave a detailed analysis for all parameters of RTV. Apart from that, $\beta$ controls the degree of smoothness. It is set a little bit large to make $\mathcal{F}$ and $\mathcal{X}$ close enough. The selection of $\mu$ depends on the processed data[25]. Therefore, parameters for the first channel of the algorithm we proposed are empirically set as Table 1, and parameters for the rest channels would be automatically generated according to strategy proposed in section 2.5.

TABLE 1  Parameters for the first channel

| Parameters | $\beta$ | $\lambda$ | $\mu$ | $\sigma$ | $\varepsilon$ | $N_{BRTV}$ |
|---|---|---|---|---|---|---|
| Numerical simulation | 0.8 | 0.003 | 6.4 | 5 | 0.01 | 4 |
| Real mouse data experiment | 0.75 | 0.0001 | 6 | 0.6 | 0.0001 | 2 |

### 3.1 Numerical simulation

The numerical simulation experiments utilize a mouse thorax phantom with 1.2% iodine contrast agent injection to validate the performance of our algorithm[36]. A polychromatic 50kVp X-ray source is employed. The spectrum is divided into eight energy channels: [16, 22), [22, 25), [25, 28), [28, 31), [31, 34), [34, 37), [37, 41), and [41, 50] keV. 640 projections are uniformly collected from fan-beam CT imaging geometry. The distances from the X-ray source to photon-counting detector (PCD) and object are 180 mm and 132 mm, respectively. The PCD consists of 512 units, and each of them covers a length of 0.1 mm. For each pixel, the physical dimension is 0.075×0.075 mm$^2$. The size of reconstructed image is 512×512×8.

### 3.1.1 Reconstruction results

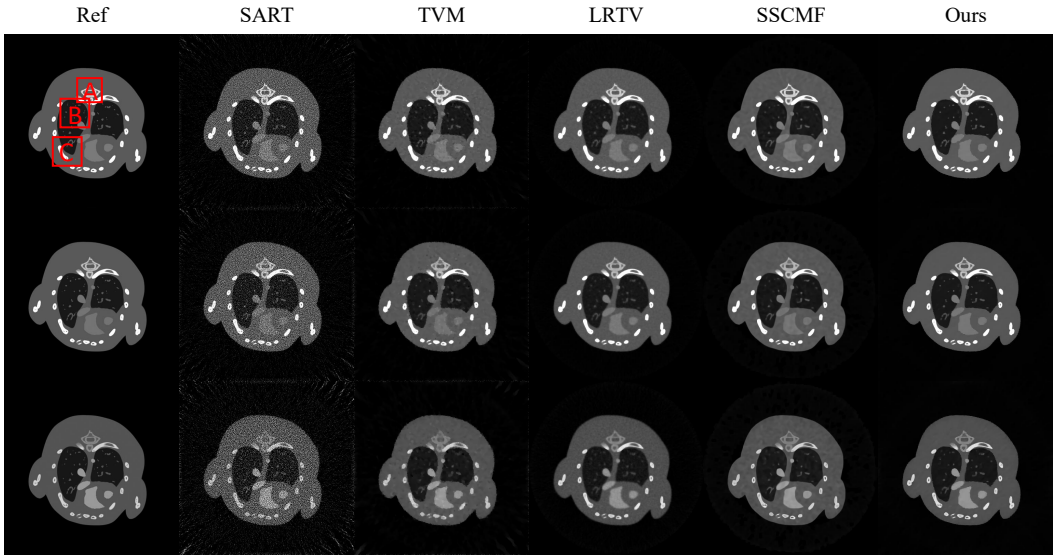

FIGURE 1  Reconstructed results of different energy channels by different algorithms. The first to third rows represent the 2$^{nd}$, 4$^{th}$, and 8$^{th}$ energy channels. The display windows for different energy channels are [0, 1.8], [0, 1.2], and [0, 0.8] cm$^{-1}$, respectively.

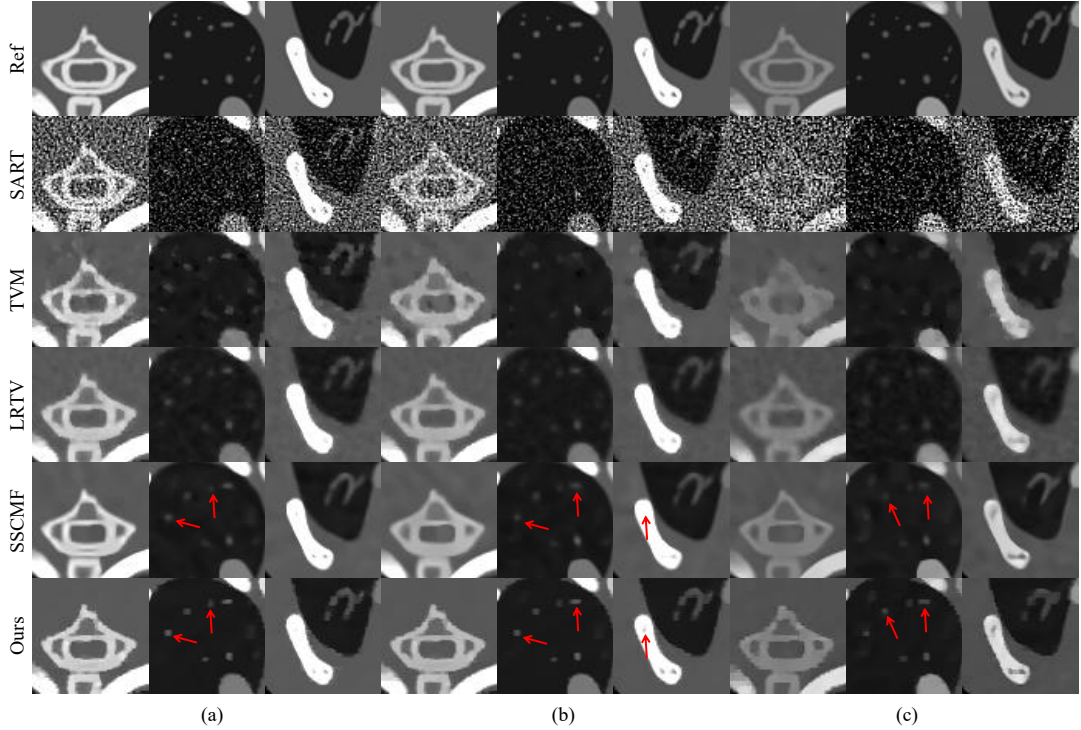

(a) (b) (c)

FIGURE 2  Magnified ROIs. (a), (b), and (c) ROIs corresponds to A, B, and C in Figure 1, respectively. The display windows for different energy channels are [0, 1.8], [0, 1.2], and [0, 0.8] cm$^{-1}$, respectively.

TABLE 2  Quantitative evaluation results of the reconstructed images

| Index | Methods | Reconstructed images (Energy channels) | | | | | | | |
|---|---|---|---|---|---|---|---|---|---|
| | | 1st | 2nd | 3rd | 4th | 5th | 6th | 7th | 8th |
| RMSE | SART | 0.3013 | 0.2122 | 0.1713 | 0.1536 | 0.1481 | 0.1519 | 0.1422 | 0.1332 |
| | TVM | 0.0708 | 0.0446 | 0.0346 | 0.0293 | 0.0264 | 0.0268 | 0.0239 | 0.0215 |
| | LRTV | 0.0628 | 0.0403 | 0.0295 | 0.0237 | 0.0211 | 0.0244 | 0.0208 | 0.0187 |
| | SSCMF | 0.0522 | 0.0310 | 0.0261 | 0.0241 | 0.0202 | 0.0193 | 0.0177 | 0.0162 |
| | Ours | **0.0510** | **0.0304** | **0.0217** | **0.0174** | **0.0158** | **0.0179** | **0.0156** | **0.0147** |
| PSNR | SART | 10.42 | 13.46 | 15.32 | 16.27 | 16.59 | 16.37 | 16.94 | 17.51 |
| | TVM | 22.99 | 27.02 | 29.21 | 30.67 | 31.56 | 31.44 | 32.44 | 33.34 |
| | LRTV | 24.04 | 27.89 | 30.62 | 32.52 | 33.51 | 32.25 | 33.64 | 34.55 |
| | SSCMF | 25.65 | 30.18 | 31.65 | 32.36 | 33.90 | 34.30 | 35.06 | 35.82 |
| | Ours | **25.84** | **30.33** | **33.26** | **35.17** | **36.05** | **34.92** | **36.13** | **36.67** |
| FSIM | SART | 0.7723 | 0.7316 | 0.7126 | 0.6889 | 0.6669 | 0.6394 | 0.6047 | 0.5965 |
| | TVM | 0.8797 | 0.8737 | 0.8729 | 0.8674 | 0.8513 | 0.8363 | 0.8290 | 0.8168 |
| | LRTV | 0.9114 | 0.8913 | 0.8908 | 0.8840 | 0.8720 | 0.8372 | 0.8325 | 0.8218 |
| | SSCMF | 0.9461 | 0.8964 | 0.8804 | 0.8697 | 0.8629 | 0.8589 | 0.8520 | 0.8430 |
| | Ours | **0.9789** | **0.9786** | **0.9783** | **0.9677** | **0.9500** | **0.9185** | **0.9191** | **0.9089** |

To compare the performance of all algorithms (SART, TVM, LRTV, SSCMF, and ours), reconstructed results in the 2nd, 4th, and 8th energy channels are demonstrated in Figure 1. On the far left column corresponds to the reference images, in which ROIs are marked by red rectangles and zoomed-in in Figure 2. The reconstructed images of SART are apparently degraded by severe noise so that some minor dots in images cannot be identified even to the naked eye. TVM reconstructed results are blurred by block artifacts, which can be clearly observed in Figure 2. Compared with the results of TVM, block artifacts are alleviated in the results of LRTV, while the details are still blurred during the reconstruction process. The phenomenon of fine structures missing and edge blurring still appears. SSCMF tends to preserve clearer and smoother edges compared with methods mentioned above while image quality in the flat regions is seriously degraded by artifacts. Besides, some important dots are smoothed by SSCMF (indicated by red rows in Figure 2). Results of the algorithm we proposed provide much finer structures and more details as well as reduction in block artifacts. Therefore, our algorithm shows an outstanding performance on elimination of noise-induced artifacts. It can both well preserve the image edge and improve the capability of finer structures and details recovery.

Table 2, listed RMSE, PSNR, and FSIM corresponding to Figure 1, demonstrates the quantitative analysis of different reconstruction algorithms. Our algorithm outdistances the other four algorithms with respective to all the evaluation measures, which means that the algorithm we proposed can obtain outstanding image qualities for all energy channels.

### 3.1.2 Material decomposition

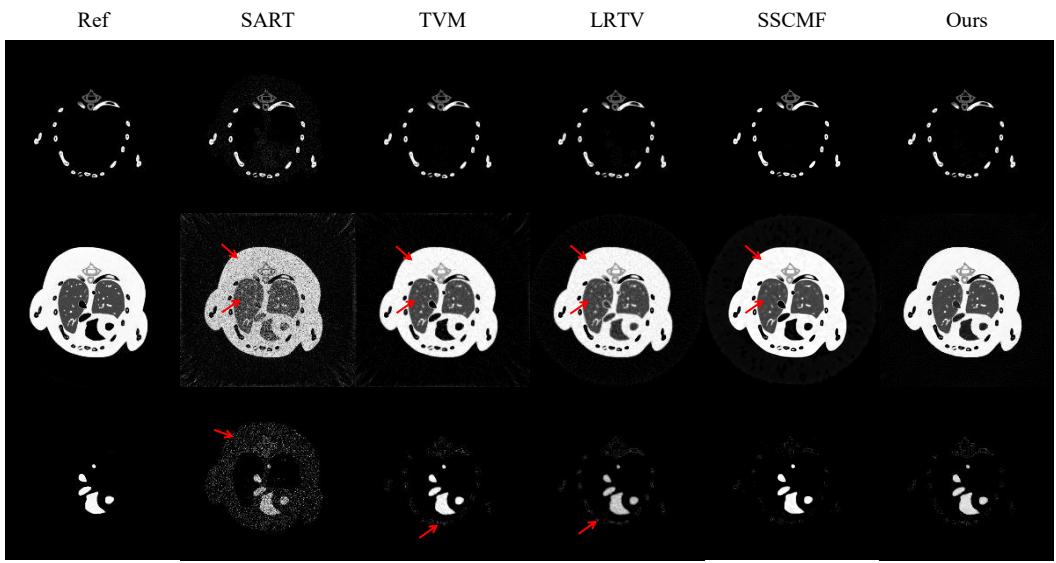

FIGURE 3  Material decomposition results of images reconstructed by different algorithms. The first to third rows represent the bone, soft tissue, and iodine contrast agent components, where the display window is [0, 1].

TABLE 3  RMSE of each decomposed basis materials in the numerical simulation study (cm$^{-1}$)

| Tissues | SART | TVM | LRTV | SSCMF | Ours |
|---|---|---|---|---|---|
| Bone | 0.0333 | 0.0117 | 0.0137 | 0.0103 | **0.0101** |
| Soft tissue | 0.1625 | 0.0435 | 0.0522 | 0.0389 | **0.0383** |
| Iodine | 0.1140 | 0.0255 | 0.0412 | **0.0191** | 0.0290 |

Figure 3 shows the decomposition results of these reconstructed images. Reconstructed spectral images were decomposed into three basis materials (bone, soft tissue and iodine contrast agent) by a classical post-reconstruction material decomposition algorithm[37]. It can be seen from the second row that the soft tissues decomposed from the results of SART, TVM,

LRTV, and SSCMF are obviously degraded to a greater or lesser degree by noise and artifacts while our algorithm presents a smoother result. However, compared with the decomposition of SSCMF, more pixels belonging to bone are wrongly decomposed as iodine contrast agent according to the result of our algorithm. The RMSE of each decomposed basis materials in the numerical simulation study are listed in Table 3, which also verifies the phenomenons mentioned above.

### 3.1.3 Convergence and computational cost

In our model, the BRTV and the nuclear norm are respectively employed to extract the inner-channel structures and characterize the inter-channel correlations of spectral CT. The convergence of our algorithm is difficult to analyze because the term of BRTV is non-convex. Thus we only numerically investigate the convergence of our algorithm. Figure 4 shows the averaged RMSE values of all energy channels vs. iteration number. The RMSEs strictly decrease with respect to the iteration number and finally converge to a stable level.

Table 4 summarizes the required time for one iteration of all algorithms. Since SART contains no regularization constraint, it has the smallest computational cost. Compared with SART, TVM and LRTV just have slightly higher computational costs due to the implementation of GPU. Because of the tensor cube matching frame in SSCMF regularization, SSCMF requires longer time than other competitors. Benefit from the parallel computation, our algorithm needs a shorter time than SSCMF.

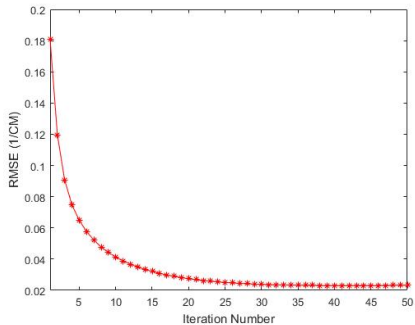

FIGURE 4    Convergence curves in terms of RMSEs.

TABLE 4    Computational costs of all reconstruction methods for one iteration (unit: s)

| Methods | SART | TVM | LRTV | SSCMF | Ours |
| --- | --- | --- | --- | --- | --- |
| Costs | 29.36 | 29.89 | 29.94 | 46.31 | 34.52 |

### 3.1.4 Ablation study

In this subsection, the ablation study is conducted to fully explore the effectiveness of two regularizers in our spectral reconstruction model. As we can see from Eq. (14), the model we proposed includes a data fidelity term, a BRTV regularizer, and a nuclear norm based regularizer. Figure 5 presents the quantitative results for ablation study. BRTV-based algorithm is applied to solve the model including a data fidelity term and a BRTV term. SVD-based algorithm represents the model integrating a data fidelity term and a nuclear norm based regularizer. Parameters of the ablation study are selected according to the PSNR maximum strategy. Results of ablation study verify that both of the two regularizers make contributions to our model. As a result, our model can effectively improve metrics in all of the eight energy channels.

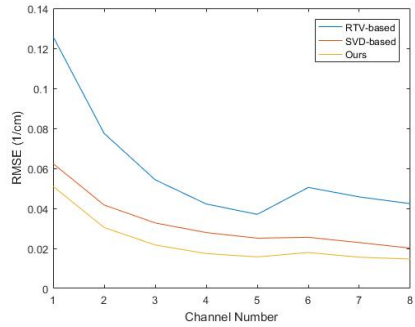

FIGURE 5    Image quality assessment for ablation study.

### 3.1.5 Adaptive weight analysis

To test the validity of multi-channel adaptive parameters generation strategy, we compare the results of our strategy with the one using same parameters in different energy channels, namely, multi-channel consistent parameters selection strategy. It is worth noting that all of the parameters of multi-channel consistent parameters selection strategy are set the same as parameters in the first channel of our strategy for a fair comparison. It can be observed from Figure 6 that multi-channel consistent parameters selection strategy tends to cause over-smoothing results especially in channels suffering from severely photon count starving. However, this problem is improved in the results by using the parameter generation strategy we proposed, which is presented in the third row of Figure 6. Images reconstructed by multi-channel adaptive parameters selection strategy have more details than multi-channel consistent parameters selection strategy.

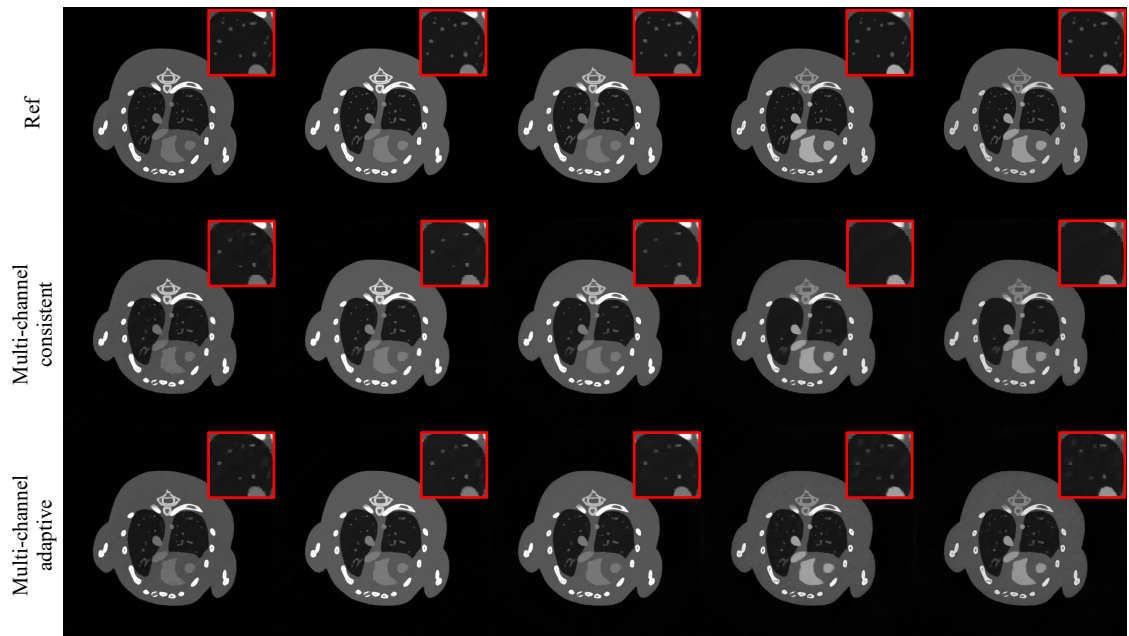

FIGURE 6    Reconstructed results with multi-channel consistent and adaptive parameters. The first to fifth columns represent the $1^{st}$, $2^{nd}$, $4^{th}$, $7^{th}$, and $8^{th}$ energy channels. The display windows for different energy channels are [0, 3], [0, 1.8], [0, 1.2], [0,0.9], and [0,0.8] cm$^{-1}$, respectively.

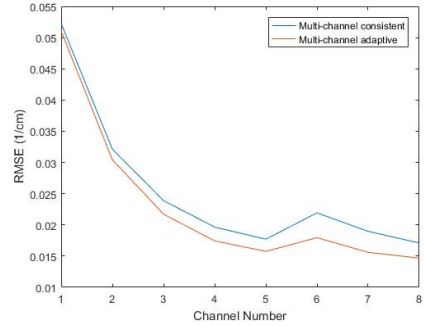

FIGURE 7　Image quality assessment with multi-channel consistent and adaptive parameters.

Figure 7 depicts the quantitative results obtained by multi-channel adaptive parameters and the competitor. It is obvious that the parameters selection strategy we proposed leads to smaller RMSEs than the case using multi-channel consistent parameters selection strategy. Both qualitative and quantitative results indicate that the proposed multi-channel adaptive parameters selection strategy shows better performance on detail preservation.

**3.2 Real mouse data experiment**

We further verify performance of the model we proposed through real mouse data in this subsection. The real mouse data was scanned by a spectral CT system developed by the Institute of High Energy Physics, Chinese Academy of Science. The distances from X-ray source to the rotational center and the detector are 200 mm and 362.8 mm, respectively. The detector consisted of 2055×60 bins with a bin size of 0.1mm×0.1mm. Projection with 540 views are evenly collected from 360°. For real spectral CT data, each channel counts the photons whose energies are greater than the threshold instead of an energy interval. Therefore, subtraction was implemented on the raw data to divide the spectrum into three energy channels: (28, 40], (40, 56], and (56, 90] keV. The image size of each channel is 512×512.

**3.2.1 Reconstruction results**

To evaluate the performance of our algorithm in real mouse data applications, reconstructed results of different algorithms are presented in Figure 8. It can be observed that the SART results are contaminated by severe noise, which blur edges and textures of the mouse image. For different channels, results of TVM present various degrees of over-smoothing and block artifacts, which means that TVM cannot adaptively suppress noise. Even though LRTV is effective in texture preservation, it shows an inability to smooth noise. SSCMF tends to cause streak artifacts especially in the local region with severe noise. This phenomenon is induced by the BM4D-part in SSCMF. Besides, some of the bones are over-smoothed during the reconstruction by SSCMF. It is to be mentioned that the streak artifacts can be eliminated by tuning parameters yet more textures of the mouse are being smoothed at the same time. These problems are improved by our algorithm, which is more effective in noise-induced artifacts elimination as well as details preservation.

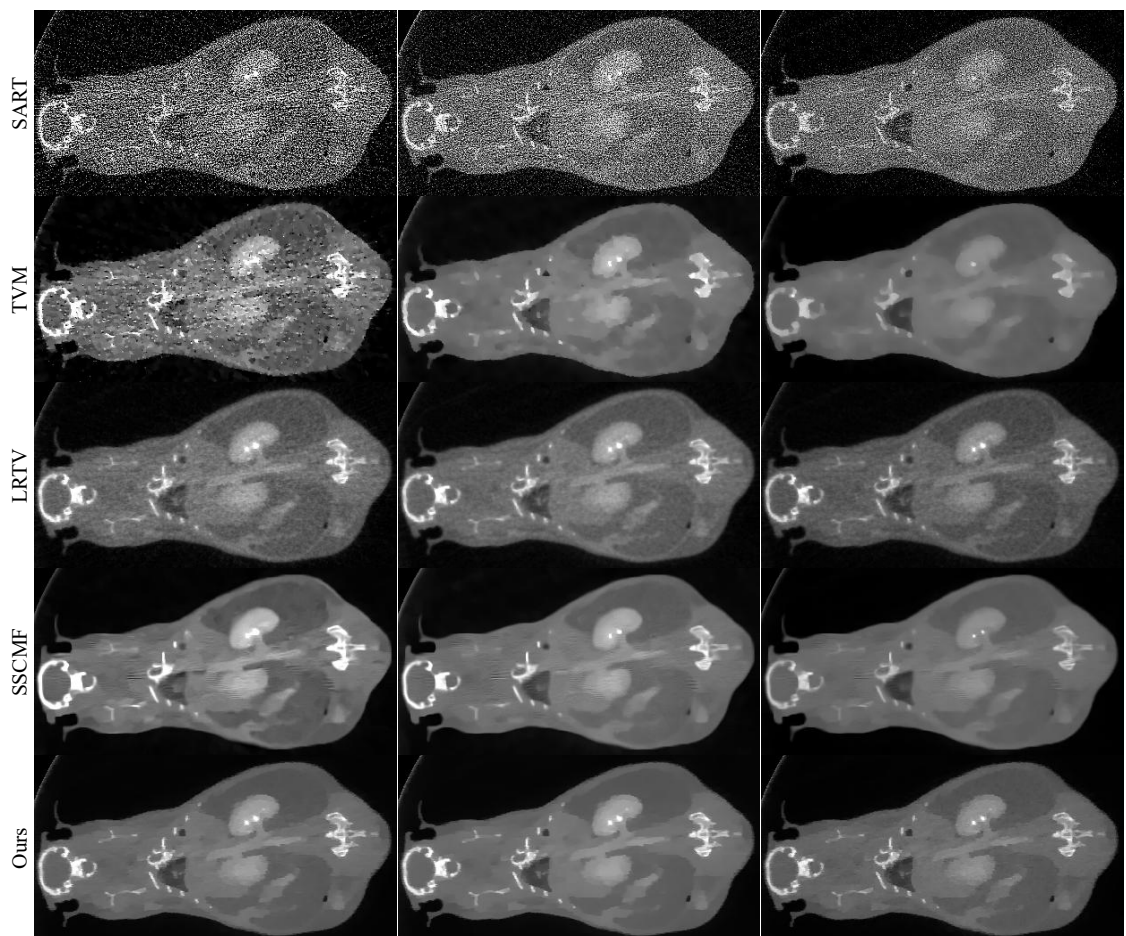

FIGURE 8   Reconstructed results of the real mouse data. The first to the last columns represent the 1$^{st}$, 2$^{nd}$, and 3$^{rd}$ energy channels. The display windows are [0, 0.5], [0, 0.6], and [0, 0.7] cm$^{-1}$, respectively.

### 3.2.2 Material decomposition

To further verify the performance of different algorithms, the material decomposition of reconstructed results from different methods are presented in Figure 9. The decompositions from reconstructed results of SART, TVM and LRTV still have obvious noise while SSCMF and our algorithm can effectively suppress noise. It is seen from the third column that many pixels from SART, TVM, and SSCMF are wrongly classified as containing iodine. LRTV and our algorithm provide more accurate iodine components. Even though our algorithm shows an relatively accurate material decomposition results compared with competing methods, some pixels from soft tissues are wrongly classified as bone from the first column in Figure 9.

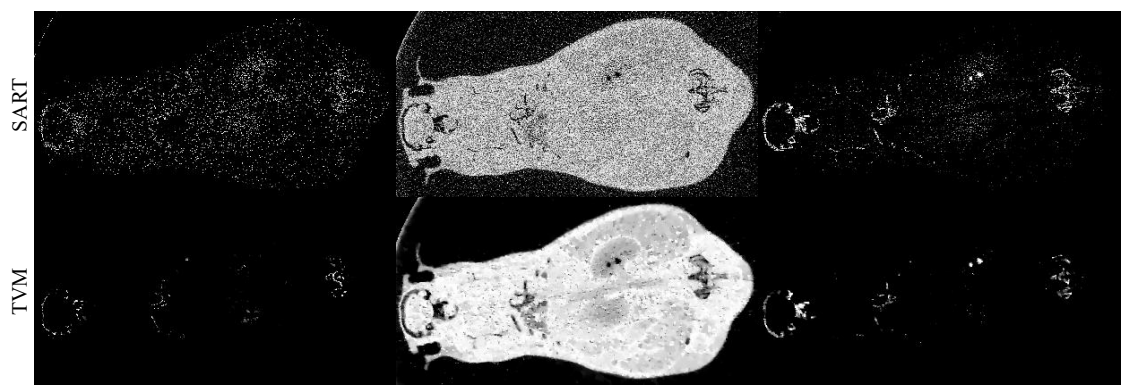

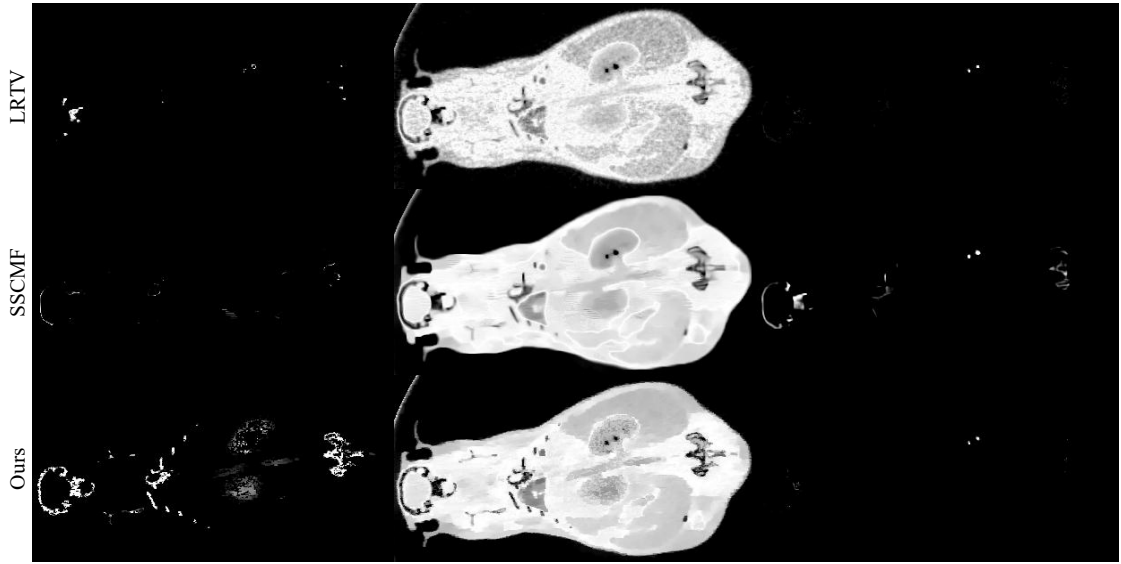

FIGURE 9   Material decomposition results of spectral images reconstructed by different algorithms. The first to third rows represent the bone, soft tissue, and iodine contrast agent components, where the display windows are [0.2, 0.5], [0.1, 1], and [0.5, 1], respectively.

## 4. Discussion and conclusion

In summary, we first propose BRTV for image restoration to address the problem that spectral CT images are severely degraded by noise and artifacts in each individual energy channel. Compared with RTV, the weights of BRTV consider both closeness to vicinity in the domain and similarity to vicinity in the range, which could simultaneously maintain edges and further reduce noise. Then we propose a spectral CT reconstruction model based on low-rank representation and structure preserving regularization. The nuclear norm and BRTV based constraints are introduced to build this spectral CT reconstruction optimization model. These two constraints are respectively utilized to characterize the inter-channel correlations and extract the inner-channel structures of spectral CT. Furthermore, we develop a split-Bregman based iterative algorithm to solve the optimization problem mentioned above. A corresponding multi-channel adaptive parameters generation strategy is proposed to dynamically generate parameters for different energy channels. Experiments on numerical simulation and real mouse datasets demonstrate that our algorithm leads to significant improvements on spectral CT reconstruction, especially achieves better results on preservation of edges and details as well as with smoother structures. Specifically, our algorithm shows an outstanding performance on elimination of noise-induced artifacts in flat area compared with competitors. At the same time, more details can be preserved by our algorithm. Besides, experiments by using the outcome of our algorithm into a classic post-reconstruction material decomposition method yields more accurate result than those of competitors.

Regarding the future work, our model has the potentiality to address other similar tensor recovery problems, such as dynamic CT reconstruction and hyperspectral image denoising. To achieve this goal, improving algorithm efficiency is the key to result in better practical value. Furthermore, more experiments will be performed to further verify the accuracy, precision, and practicability of the proposed algorithm in the future.

**APPENDIX A**

$$\mathcal{X}^{k+1} = \arg\min_{\mathcal{X}} \sum_{s=1}^{S} \left( \|AX_s - Y_s\|_2^2 + \lambda \sum_p BRTV(X_s(p)) \right) + \beta \|\mathcal{X} - \mathcal{F}^k - \mathcal{V}^k\|_F^2, \tag{A1}$$

Eq. (A1) can be rewritten as follows:

$$\arg\min_{\{X_s\}_{s=1}^S} \sum_{s=1}^{S} \left( \|AX_s - Y_s\|_2^2 + \lambda \sum_p BRTV(X_s(p)) + \beta \|X_s - F_s^k - V_s^k\|_F^2 \right), \tag{A2}$$

where $F_s^k \in \mathcal{R}^{N_1 \times N_2}$ and $V_s^k \in \mathcal{R}^{N_1 \times N_2}$ are the $s^{th}$ channel images of tensors $\mathcal{F}^k$ and $\mathcal{V}^k$, respectively. Alternatively, Eq. (A2) is equivalent to solve the following objective function:

$$\arg\min_{\{X_s\}_{s=1}^S} \|AX_s - Y_s\|_2^2 + \lambda \sum_p BRTV(X_s(p)) + \beta \|X_s - F_s^k - V_s^k\|_F^2, \tag{A3}$$

We linearize $\|AX_s - Y_s\|_2^2$ at point $X_s^k$. Eq. (A3) can be approximately expressed as:

$$\arg\min_{\{X_s\}_{s=1}^S} \|AX_s^k - Y_s\|_2^2 + 2\langle A^T(AX_s^k - Y_s), X_s - X_s^k \rangle + \gamma \|X_s - X_s^k\|_2^2 + R(X_s), \tag{A4}$$

$$R(X_s) = \lambda \sum_p BRTV(X_s(p)) + \beta \|X_s - F_s^k - V_s^k\|_F^2. \tag{A5}$$

A more concise form of Eq. (A4) can be reformulated as follows:

$$\arg\min_{\{X_s\}_{s=1}^S} \|X_s - X_s^{k+1/2}\|_2^2 + \frac{1}{\gamma} R(X_s), \tag{A6}$$

$$X_s^{k+1/2} = X_s^k - \frac{1}{\gamma} A^T(AX_s^k - Y_s), \tag{A7}$$

where $X_s^{k+1/2}$ is an intermediate image. Eq. (A7) can be regarded as the simultaneous algebraic reconstruction technique (SART), in which $\gamma$ can be omitted[34]. To solve Eq. (A6), we first discuss the x-direction measure of BRTV. It can be written as:

$$\sum_p \frac{D_x(p)}{\widetilde{L}_x(p) + \varepsilon} = \sum_p \frac{\sum_{q \in R(p)} k_{p,q} \cdot |(\partial_x X_s)_q|}{\left|\sum_{q \in R(p)} h_{p,q} \cdot (\partial_x X_s)_q\right| + \varepsilon} = \sum_q \sum_{p \in R(q)} \frac{k_{p,q}}{\left|\sum_{q \in R(p)} h_{p,q} \cdot (\partial_x X_s)_q\right| + \varepsilon} |(\partial_x X_s)_q|$$

$$\approx \sum_q \sum_{p \in R(q)} \frac{k_{p,q}}{\widetilde{L}_x(p) + \varepsilon} \frac{1}{|(\partial_x X_s)_q| + \varepsilon_{X_s}} (\partial_x X_s)_q^2 = \sum_q \widetilde{u}_{xq} w_{xq} (\partial_x X_s)_q^2. \tag{A8}$$

$\varepsilon_{X_s}$ is introduced for numerical stability. $\widetilde{u}_{xq}$ and $w_{xq}$ are set as:

$$\widetilde{u}_{xq} = \sum_{p \in R(q)} \frac{k_{p,q}}{\widetilde{L}_x(p) + \varepsilon}, \quad w_{xq} = \frac{1}{|(\partial_x g)_q| + \varepsilon_{X_s}}. \tag{A9}$$

A similar process can be used to the y-direction measure in BRTV. Therefore, Eq. (A6) can be written in a matrix form by plugging Eq. (A5):

$$\arg\min_{\{X_s\}_{s=1}^S} \left(v_{X_s} - v_{X_s^{k+1/2}}\right)^T \left(v_{X_s} - v_{X_s^{k+1/2}}\right) + \frac{\lambda}{\gamma}\left(v_{X_s}^T C_x^T \widetilde{U}_x W_x C_x v_{X_s} + v_{X_s}^T C_y^T \widetilde{U}_y W_y C_y v_{X_s}\right)$$
$$+ \frac{\beta}{\gamma}\left(v_{X_s} - v_{F_s^k} - v_{V_s^k}\right)^T \left(v_{X_s} - v_{F_s^k} - v_{V_s^k}\right), \quad (A10)$$

where $v_{X_s}$, $v_{X_s^{k+1/2}}$, $v_{F_s^k}$, and $v_{V_s^k}$ represent the vectorized images of $X_s$, $X_s^{k+1/2}$, $F_s^k$, and $V_s^k$, respectively. $C_x$ and $C_y$ are matrices of the forward difference operator. $\widetilde{U}_x$, $\widetilde{U}_y$, $W_x$, and $W_y$ are the diagonal matrices consisting of $\widetilde{u}_{xq}$, $\widetilde{u}_{yq}$, $w_{xq}$, and $w_{yq}$, respectively. For simplicity, we let $\beta_1 = \frac{\beta}{\gamma}$ and $\lambda_1 = \frac{\lambda}{\gamma}$. Therefore, the $\mathcal{X}^{k+1}$ can be updated using the following iterative equation:

$$\left[(1+\beta_1)I + \lambda_1\left(C_x^T \widetilde{U}_x^t W_x^t C_x + C_y^T \widetilde{U}_y^t W_y^t C_y\right)\right] \cdot v_{X_s^{k+1}}^{t+1} = v_{X_s^{k+1/2}} + \beta_1\left(v_{F_s^k} + v_{V_s^k}\right), \quad (A11)$$

where $1 \leq t \leq N_{BRTV}$ represents the $t^{th}$ iteration of BRTV during each round of the iterative reconstruction. It can be easily proved that $(1+\beta_1)I + \lambda_1\left(C_x^T \widetilde{U}_x^t W_x^t C_x + C_y^T \widetilde{U}_y^t W_y^t C_y\right)$ is symmetric positive, which means that a unique solution of (A11) exists. The preconditioned conjugate gradient method is utilized to solve (A11).

**APPENDIX B**

$$\mathcal{F}^{k+1} = \arg\min_{\mathcal{F}} \mu\|\mathcal{F}\|_* + \beta\|\mathcal{X}^{k+1} - \mathcal{F} - \mathcal{V}^k\|_F^2, \quad (B1)$$

The sub-problem (B1) is a normal problem of nuclear norm, which can be solved by the singular value thresholding algorithm[38]. Consider the singular value decomposition of a matrix $unfold(\mathcal{X}^{k+1} - \mathcal{V}^k) \in \mathcal{R}^{S \times (N_1 \times N_2)}$ with rank $r$:

$$unfold(\mathcal{X}^{k+1} - \mathcal{V}^k) = U \Sigma V^*, \quad \Sigma = diag(\{\vartheta_i\}_{1 \leq i \leq r}), \quad (B2)$$

where $U \in R^{S \times r}$ and $V^* \in R^{r \times (N_1 \times N_2)}$ are matrices with orthonormal columns. $\vartheta_i$ is the $i^{th}$ singular value of $unfold(\mathcal{X}^{k+1} - \mathcal{V}^k)$. We set $\mu_1 = \frac{\mu}{2\beta}$. Eq. (B1) can be solved as follows:

$$\mathcal{F}^{k+1} = fold(U\mathcal{D}_{\mu_1}(\Sigma)V^*), \quad \mathcal{D}_{\mu_1}(\Sigma) = diag(\{\max(\vartheta_i - \mu_1, 0)\}). \quad (B3)$$

**ACKNOWLEDGMENTS**
This work was supported by the National Natural Science Foundation of China under Grant 61771003 and 12175026; Graduate Research and Innovation Foundation of Chongqing, China under Grant CYB21044.

**COFLICT OF INTEREST**

The authors have no conflict of interest to disclose.